\documentclass[twocolumn]{aastex7}
\newcommand*{\Euclid}{\textit{Euclid} }
\newcommand{\sfont}[1]{{\scriptscriptstyle\rm #1}}
\newcommand{\IE}{\ensuremath{I_\sfont{E}}\,}
\newcommand{\YE}{\ensuremath{Y_\sfont{E}}\,}
\newcommand{\JE}{\ensuremath{J_\sfont{E}}\,}
\newcommand{\HE}{\ensuremath{H_\sfont{E}}\,}

\begin{document}

\title{Euclid Quick Data Release (Q1) - Spectroscopic search, classification and analysis of ultracool dwarfs in the Deep Fields}

\author[gname=Carlos,sname=Dominguez-Tagle]{C. Dominguez-Tagle}
\affiliation{Instituto de Astrofisica de Canarias, La Laguna, Spain}
\affiliation{Universidad de La Laguna, Dpto. Astrofisica, La Laguna, Spain}
\email{carlos.dominguez@iac.es}

\author[gname=Marusa,sname={\v Z}erjal]{M. {\v Z}erjal}
\affiliation{Instituto de Astrofisica de Canarias, La Laguna, Spain}
\affiliation{Universidad de La Laguna, Dpto. Astrofisica, La Laguna, Spain}
\email{maruska.zerjal@iac.es}

\author[gname=Nafise,sname=Sedighi]{N. Sedighi}
\affiliation{Instituto de Astrofisica de Canarias, La Laguna, Spain}
\affiliation{Universidad de La Laguna, Dpto. Astrofisica, La Laguna, Spain}
\email{nafise.sedighi@iac.es}

\author[gname=Pedro,sname=Mas-Buitrago]{P. Mas-Buitrago}
\affiliation{Centro de Astrobiologia (INTA-CSIC),Villanueva de la Cañada, Madrid, Spain}
\email{pmas@cab.inta-csic.es}

\author[gname=Eduardo,sname=Martin]{E. L. Martin}
\affiliation{Instituto de Astrofisica de Canarias, La Laguna, Spain}
\affiliation{Universidad de La Laguna, Dpto. Astrofisica, La Laguna, Spain}
\email{ege@iac.es}

\author[gname=Junyan,sname=Zhang]{J.-Y. Zhang}
\affiliation{Instituto de Astrofisica de Canarias, La Laguna, Spain}
\affiliation{Universidad de La Laguna, Dpto. Astrofisica, La Laguna, Spain}
\email{junyan.zhang@iac.es}

\author[gname=Nikola,sname=Vitas]{N. Vitas}
\affiliation{Instituto de Astrofisica de Canarias, La Laguna, Spain}
\affiliation{Universidad de La Laguna, Dpto. Astrofisica, La Laguna, Spain}
\email{nikola.vitas.iac@gmail.com}

\author[gname=Victor,sname=Bejar]{V. J. S. B{\' e}jar}
\affiliation{Instituto de Astrofisica de Canarias, La Laguna, Spain}
\affiliation{Universidad de La Laguna, Dpto. Astrofisica, La Laguna, Spain}
\email{victor.bejar@iac.es}

\author[gname=Styliani,sname=Tsilia]{S. Tsilia}
\affiliation{Instituto de Astrofisica de Canarias, La Laguna, Spain}
\affiliation{Universidad de La Laguna, Dpto. Astrofisica, La Laguna, Spain}
\email{stella@iac.es}

\author[gname=Sara,sname=Muñoz Torres]{S. Muñoz Torres}
\affiliation{Instituto de Astrofisica de Canarias, La Laguna, Spain}
\affiliation{Universidad de La Laguna, Dpto. Astrofisica, La Laguna, Spain}
\email{sara.munoz@iac.es}

\author[gname=Nicolas,sname=Lodieu]{N. Lodieu}
\affiliation{Instituto de Astrofisica de Canarias, La Laguna, Spain}
\affiliation{Universidad de La Laguna, Dpto. Astrofisica, La Laguna, Spain}
\email{nlodieu@iac.es}

\author[gname=David,sname=Barrado]{D. Barrado}
\affiliation{Centro de Astrobiologia (INTA-CSIC),Villanueva de la Cañada, Madrid, Spain}
\email{barrado@cab.inta-csic.es}

\author[gname=Enrique,sname=Solano]{E. Solano}
\affiliation{Centro de Astrobiologia (INTA-CSIC),Villanueva de la Cañada, Madrid, Spain}
\email{esm@cab.inta-csic.es}

\author[gname=Patricia,sname=Cruz]{P. Cruz}
\affiliation{Centro de Astrobiologia (INTA-CSIC),Villanueva de la Cañada, Madrid, Spain}
\email{pcruz@cab.inta-csic.es}

\author[gname=Ramarao,sname=Tata]{R. Tata}
\affiliation{Department of Physics and Astronomy, Ohio University, USA}
\email{tatar@ohio.edu}

\author[gname=Phan,sname=Phan-Bao]{N. Phan-Bao}
\affiliation{Department of Physics, International University, Ho Chi Minh City, Vietnam}
\affiliation{Vietnam National University, Ho Chi Minh City, Vietnam}
\email{pbngoc@hcmiu.edu.vn}

\author[gname=Adam,sname=Burgasser]{A.~Burgasser}
\affiliation{Department of Astronomy \& Astrophysics, UC San Diego, USA}
\email{aburgasser@ucsd.edu}


\begin{abstract}
The Near-Infrared Spectrometer and Photometer onboard the \Euclid space mission has obtained near-infrared (NIR) slitless spectra of millions  of objects, including hundreds of ultracool dwarfs.
\Euclid observations retrieve images and spectra simultaneously. This observing mode marks a new era in the discovery of new objects, such as L- and T-type dwarfs, which can be found from direct identification through the H$_2$O and CH$_4 $ absorption bands. NISP spectral resolution ($R \sim $~450) is enough to classify the objects by the spectral type using known standard templates.
Q1  provided more than 4 million NIR spectra in one visit to the Euclid Deep Fields. 
The large amount of spectra released in these fields allowed us to: 
a) confirm the ultracool dwarf nature   of almost half of the  photometric candidates    compiled   by \cite{ZANG2024}; 
b) discover at least  11  new late L- and T-type dwarfs by a specific spectral index search in Q1 data;  and 
c) spectroscopically confirm one hundred more candidates from a new photometric selection conducted by {\v Z}erjal et al. (in prep.).   We present a preliminary list of \Euclid  ultracool dwarf  templates built by the combination of the best spectra from all these searches.
We include the first spectral analysis of  confirmed  ultracool dwarfs  from Q1 data;  spectral classifications; determination of effective temperatures;  H$_2$O, CH$_4 $ and  NH$_3 $ spectral indices; and measurements of the K{\small{I}}~absorption doublet. This paper is a first step in the study of \Euclid   ultracool dwarfs    and will be improved with each subsequent data release.

\end{abstract}





\section{\label{sc:Intro}Introduction}
The advent of the \Euclid space mission \citep{EuclidSkyOverview} is expected to foster major advancements  in many fields of astrophysics.  The \Euclid Quick Data Release (Q1) includes slitless near-infrared (NIR) spectra of 4.3 million objects observed with the Near-Infrared Spectrometer and Photometer  \citep[NISP;][]{EuclidSkyNISP}. Among those millions of sources, several hundred are   ultracool dwarfs (UCDs),   which have effective temperatures below 2700~K \citep{Kirkpatrick95} and down to $\sim$~250~K for the coldest
object observed to date \citep{Luhman14, Luhman24}. 
The spectral classification system for UCDs follows 
their decreasing temperatures:  late M-type (M7 and later, \citealt{Kirkpatrick97},  L-type \citep{ege97,ege98,ege99,Kirkpatrick99,Kirkpatrick00},  T-type \citep{Burgasser02, Geballe02, Burgasser06}, and  Y-type dwarfs \citep{Delorme08,Cushing11, Kirkpatrick11}.

This paper focuses on the spectroscopic capabilities of \Euclid in the field of  ultracool dwarf (UCD)    science, and is part of a collection of papers dedicated to the study of  UCDs.  Here we investigate whether NISP's spectral resolution and sensitivity are sufficient to analyze the molecular bands 
of H$_2$O, CH$_4$ and  NH$_3$ (see \citealt{EGE2021amonia}), as well as other spectral features, in order to classify them and to study their physical properties. To this end, we focus on four objectives:
\begin{itemize}
    \item To  confirm     the    ultracool dwarf nature of the   photometric candidates published by \cite{ZANG2024}, through the analysis of \Euclid spectra.
    \item To discover new late-L and T-type UCDs by a specific spectral index search in the full Q1 database.
    \item To  confirm the photometric UCD candidates selected from Q1 data by {\v Z}erjal et al. (in prep.).
    \item To classify the UCD candidates by spectral type and to investigate the determination of effective temperatures, spectral indices, pseudo-equivalent widths of the K{\small{I}}~absorption doublet, and the possibility of identifying objects with unusual high radial velocities.
\end{itemize}

\begin{figure}
\includegraphics[width=\hsize]{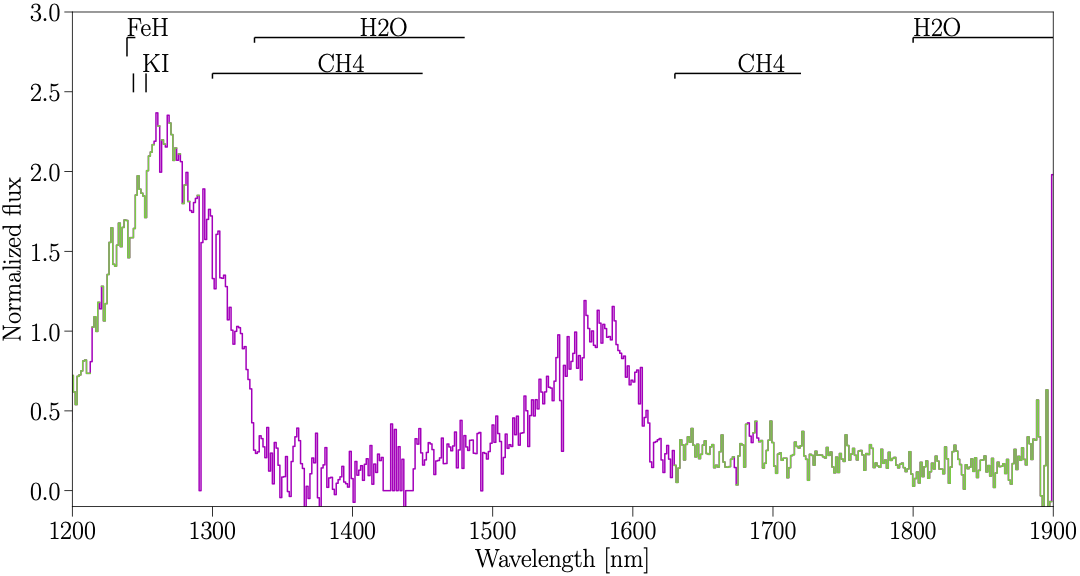} 
   \caption{Spectrum of E597913, a T7 object from  Zhang's compilation with spectra available in Q1 data. The flux is normalized at 1600~nm. The spectrum is plotted without any smoothing. The green lines represent the spectra with \texttt{NDITH}~$>$~2 and the  magenta lines  indicate the values with  \texttt{QUALITY}~$<$~0.7.}
    \label{fig.example}
\end{figure}

\Euclid observations retrieve visible and NIR images and NIR spectra simultaneously, using its two instruments VIS \citep{EuclidSkyVIS} and NISP in a Reference Observing Sequence  \cite[ROS;][]{Scaramella-EP1}. Each ROS includes four images in the four bands
(\IE,  \YE, \JE, and \HE) and four spectra by performing a dithering pattern. The four dithering positions (Level-1 data) can be coadded to increase the signal-to-noise ratio (SNR) and to eliminate unwanted artifacts (cosmic rays, bad pixels, optical ghosts, etc.). For most of the \Euclid survey, only one ROS is planned, with the exception of the Euclid Deep Fields (EDFs), which will be observed repeatedly during the mission. These deep fields are grouped in three regions North, Fornax and South: EDF-N, EDF-F, and EDF-S, respectively.

Q1 \citep{Q1-TP001} provides calibrated data of one visit to the EDFs. The data include coadded and calibrated images and extracted spectra, as well as catalogs for the different type of objects (stars, galaxies, etc.). 
Q1 data include 4.3 million extracted NISP spectra which cover the wavelength range 1200 to 1900~nm \citep[see][]{Q1-TP006,Q1-TP007}. The spectra files include wavelength (in air), flux density, the flux density variance, along with the following flags~\footnote{\url{https://euclid.esac.esa.int/dr/q1/dpdd/}}:
\begin{description}
    \item[\texttt{NDITH}] number of individual spectra combined for this pixel.
    \item[\texttt{MASK}] a bit mask for the current pixel signaling saturation, ghost, persistence, etc. The specific bit meaning is included in the header of the DQ extension of each SIR spectrogram.
    \item[\texttt{QUALITY}] a number between 0 and 1 that represents the ratio between the weighted sum of the pixels combined and the weighted sum of all the pixels. 
     
\end{description}

While slitless spectra observing mode provides the spectrum of every object in the field of view, there is a significant probability of contamination among the objects. In addition, given that NISP has a mosaic of 16 detectors with gaps between
them, all or part of the object spectrum could fall into these gaps for some of the dithering positions (out of four in a single ROS). An example of a Q1 spectrum of E597913 (alias composed by the first 6 digits of the \Euclid  ID) of a confirmed T7 UCD in the EDF-N identified by \cite{ZANG2024} is shown in Fig.~\ref{fig.example}. The  spectrum is displayed in two colors: green lines represent the spectra with \texttt{NDITH}~$>$~2 and  magenta lines  represent the spectra with  \texttt{QUALITY}~$<$~0.7. The number of dithering positions can change with  wavelength, so plotting the spectrum in two colors helps to identify any unreliable portions (i.e. from unidentified contamination). We chose the value of 0.7 for  \texttt{QUALITY} as it filters most of the artifacts.
Future data releases will be based on the combination of more dithering positions, so that \texttt{QUALITY} and \texttt{NDITH} values will increase.

This paper is organized as follows:
Sect.~\ref{sc:data} describes the photometric candidates and the non-photometric search of late L- and T-type objects in Q1 data;
Sect.~\ref{sc:confirm} describes the spectroscopic confirmation and classification;
Sect.~\ref{sc:template} shows the preliminary {\it Euclid} UCD templates;
Sect.~\ref{sc:anal} reviews the spectral analysis of the confirmed UCDs;
and Sect.~\ref{sc:concl} presents our conclusions.

\section{\label{sc:data}Retrieving UCD candidates from Q1 data}

We have selected UCD candidates from a previously published photometric catalog and from an independent spectral index search.  
The  previous photometric catalog is taken from \citet[hereafter Zhang's compilation]{ZANG2024}. It is a selection of known objects filtered to find UCD candidates. The ones spectroscopically confirmed as UCDs by this work are called hereafter ``UCD benchmarks''.
Our independent spectral index search of late L- and T-type objects in the 4.3 million Q1 spectra adds some more UCDs.
And, a new photometric catalog is currently in preparation using the UCD benchmarks and the feedback from this work. Its is called hereafter {\v Z}erjal's catalog ({\v Z}erjal et al., in prep.).

\begin{figure}
   \centering
   \includegraphics[width=1\hsize]{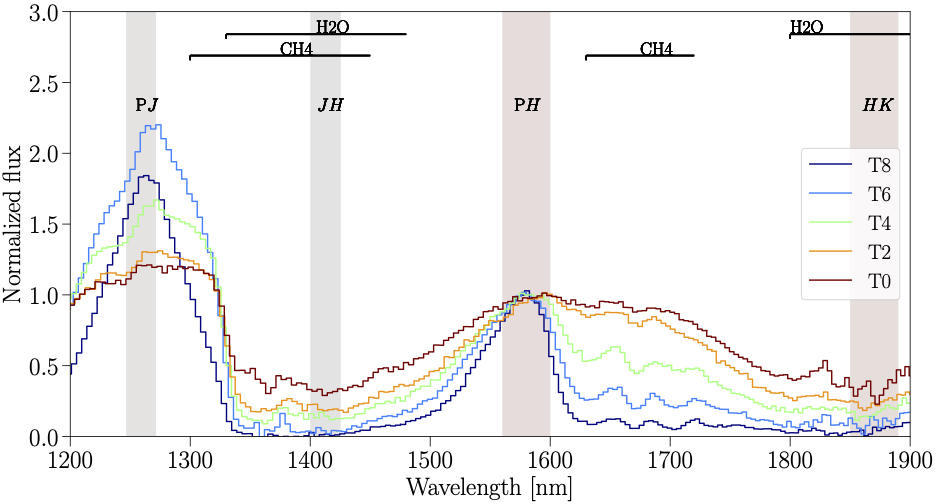} 
   \caption{Spectral regions used for the spectral index search,  plotted over the spectra of the T-dwarf standard templates (references in Appendix~\ref{sc.splat}).   }
   \label{fig.indices}
\end{figure}

\subsection{\label{sc:dataJerry} Zhang's compilation }

In preparation for {\it Euclid} observations,  \cite{ZANG2024} performed a photometric selection of UCDs in the region of the 
EDF-N, and selected candidates from the catalogs published by \cite{Reyle2018} and  \cite{Carnero2019} for the  EDF-F and EDF-S. Zhang's compilation includes  a total of 515 candidates. 

They selected eight candidates which they  confirmed as UCDs by obtaining reconnaissance spectra from the 8.2-m Very Large Telescope (VLT) and 10.4-m Gran Telescopio Canarias (GTC). Five of the objects from this selection are found in Q1 data (our alias in parentheses): WISEA J035909.75$-$474056.8 (E597913), WISEA J035231.80$-$491059.4 (E581332), WISEA J033234.35$-$273333.8 (E531434), WISEA J040254.85$-$470440.9 (E607287), and WISEA J175730.71+674138.4 (E269377).  

Zhang's compilation includes NIR photometry from the Two Micron All Sky Survey \citep[2MASS;][]{2003tmc..book.....C,2006AJ....131.1163S} for all the objects in the EDF-N, and for a few ones in the EDF-F and EDF-S.  The latter two fields have most of their photometry  from the VISTA Hemisphere Survey \citep[VHS;][]{VHS2013}. The use of different sources produces a difference in depth between the sample for EDF-N and that for the other two fields, as we will see in Sect.~\ref{sc:Jerryspec}.

 In this work we label these photometric measurements without any suffix (e.g.,  $J$ )  to differentiate from NISP photometry (\YE, \JE, and \HE ).

\begin{table}[h!]
\caption{\label{tab.directshort} New late-L and T dwarfs found by the spectral index search  algorithm.}
\centering

\begin{tabular}{llrrr}
\hline\hline
Alias & SpT & P$J$-$JH$ & P$H$-$JH$ & P$H$-$HK$ \\
\hline
E528241 &  T5  & 0.9 & 0.7 & 0.7 \\
E265716 &  T4p & 0.9 & 0.9 & 0.9 \\
E523574 &  T4  & 0.9 & 0.8 & 0.8 \\
E273062 &  T3p  & 0.9 & 0.8 & 0.6 \\
E536416 &  T2:  & 0.8 & 0.8 & 0.8 \\
E273015 &  T1:  & 0.8 & 0.8 & 0.8 \\
E644720 &  T1: & 0.7 & 0.6 & 0.6 \\
E511520 &  T1:  & 0.8 & 0.8 & 0.9 \\
E267056 &  L9:  & 0.7 & 0.7 & 0.5 \\
E517518 &  L9  & 0.6 & 0.6 & 0.5 \\
E271934 &  L9p & 0.7 & 0.6 & 0.4 \\
\hline
\end{tabular}
\tablecomments{  The third to fifth  columns are the search indices defined in the text.  The spectral type (SpT) is described in Sect.~\ref{sc:resu}.  }
\end{table}

\subsection{\label{sc:direct}Spectral index search  for late-L and T dwarfs }
Given that  Zhang's compilation had a very  low number of  late L- and T-type candidates, and that Q1 data released 4.3 million spectra, we decided to search for the accentuated spectral features of late-L and T dwarfs  directly in the spectra database without any photometric analysis.
In particular, the combined absorption of H$_2$O and CH$_4 $  is so evident in the spectral range covered by NISP that a fast algorithm computing ratios can easily detect these absorption bands. We developed an algorithm that filters the data to eliminate unreliable values using \texttt{QUALITY} and \texttt{NDITH} flags  and computes three ratios from four narrow spectral regions. The values for these regions were adapted from the NIR spectral indices defined by \cite{Burgasser06}, to the expected \Euclid spectra of T dwarfs. These regions are plotted together with four T-dwarf standard templates in Fig.~\ref{fig.indices}.
The regions are labeled as P$J$ and  P$H$ for the peak regions, and  $JH$ and $HK$ for   the    deepest region of the water absorption bands. We  selected these regions because  \Euclid observations are free from telluric absorption bands. The search algorithm normalizes the integrated fluxes over the regions to the peak values and computes three ratios P$J$-$JH$,  P$H$-$JH$, and P$H$-$HK$. It works in two steps, first with  the spectra free from unreliable values, then for the spectra convolved with a gaussian profile (boxcar = 7). In both cases it filters by the minimum values of 0.6 and 0.5 for P$J$-$JH$ and  P$H$-$JH$, respectively. In the second step it filters by P$H$-$HK$ $>$0.35. We calibrated these values running the search algorithm on the spectra of known T dwarfs.
We ran the search algorithm  over the  entire database with satisfying results. The output included numerous spectra filled with contamination from other sources that resembled portions of expected spectral energy distribution. After a visual inspection, we kept the best findings, which turned into  11  new UCDs not published before. Table~\ref{tab.directshort} lists these  11  objects  discovered by the search algorithm, and gives the values of the search indices.  This table includes their spectral type, described in Sect.~\ref{sc:class}. It is interesting to see that the  11  objects spectral types span homogeneously from L9 to T5. The objects retrieved by the algorithm that were   previously known,   are described in Sect.~\ref{sc:resu}.   We also conducted a search for Y dwarfs, with no concluding results.

\subsection{\label{sc:dataMarusa}{\v Z}erjal's catalog  }

 A new catalog of photometric UCD candidates from Q1 is under preparation by {\v Z}erjal et al. (in prep.). Special attention has been given to the selection of the point-source objects, as the vast majority of \Euclid detections are extragalactic. The first step is focused on isolating point sources based on morphological criteria and relatively strict data quality cuts for each object (SNR $>$ 3 in the \IE, \YE, and \HE bands). These cuts reduce the sample to just under 2\% of the original Q1 dataset.  UCD candidates are then selected using a color–color diagram (\YE - \HE vs. \IE - \YE). The inclusion of the optical band is crucial, as it allows UCDs, especially T dwarfs, to stand out from the rest of the sample. Guided by the UCD benchmarks, the selection criterion is set to  \IE - \YE $>$ 2.5, a typical color for late-M dwarfs. To date, this selection has yielded more than 5000 UCD candidates, some of which are included in this paper.

\section{\label{sc:confirm}Spectroscopic confirmation and classification }

\subsection{\label{sc:class}Spectral type (SpT) classification  }
\begin{figure}
\includegraphics[width=\hsize]{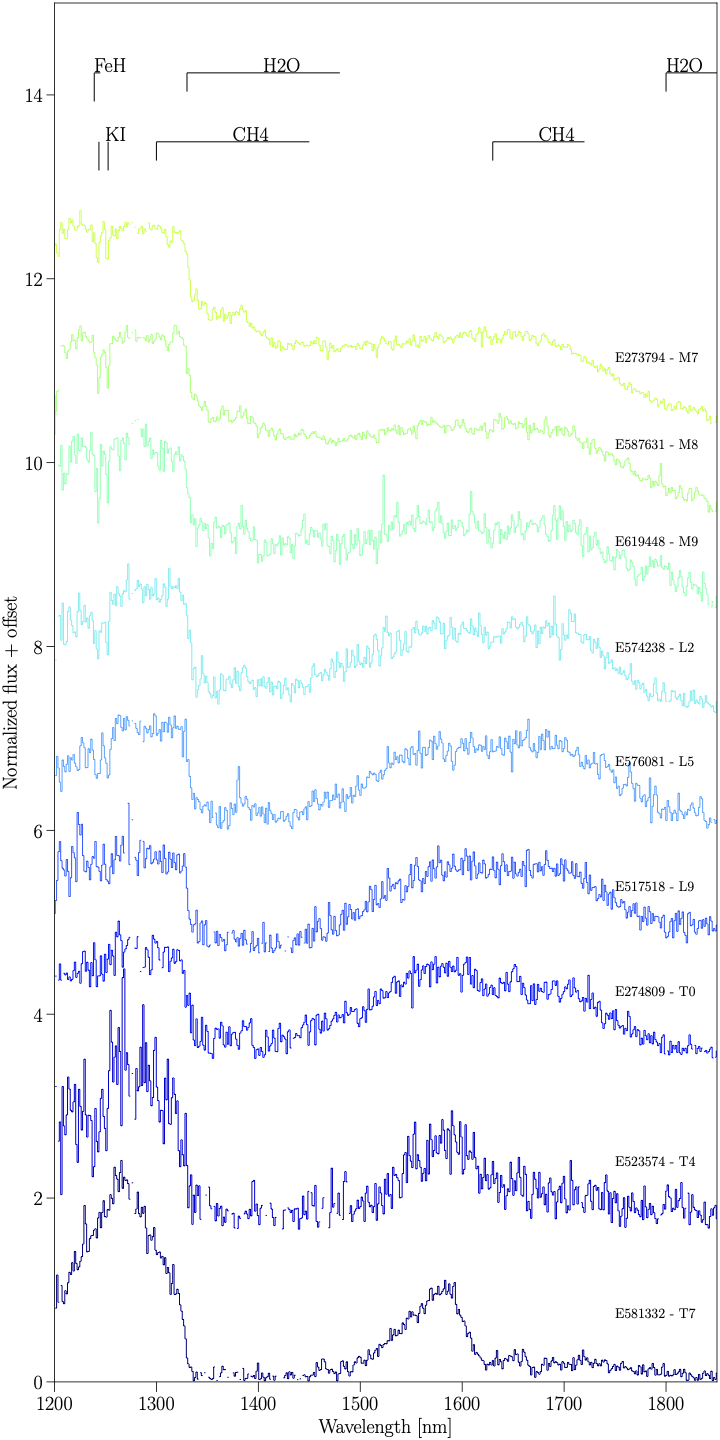} 
\caption{Spectral sequence for objects from M7 to T7 that shows {\it Euclid's} ability to  classify  UCDs.  The flux is normalized at 1600~nm. The spectra are plotted without any smoothing. The missing values are due to  artifact removal. }
\label{fig.SpectralSequence}
\end{figure}

We classified the objects by  comparing the spectra to the standard templates compiled in SPLAT \citep{SPLAT}  and selecting the  spectral type by the best fit.
Appendix~\ref{sc.splat} gives the details of the  standard templates. 
Owing to the reduced wavelength range and the presence of artifacts, the best fit is obtained by two estimation methods.
The first one is by minimizing chi-squared values over the full wavelength range. The second method is  by residual minimization over four wavelength ranges  defined by NISP wavelength range and the telluric absorption bands present in the  standard templates: 1210$-$1350~nm, 1350$-$1530~nm, 1530$-$1800~nm,  and 1800$-$1880~nm. 
The results are then weighted by the value of  \texttt{QUALITY} and \texttt{NDITH}  parameters, described in Sect.~\ref{sc:Intro}.
 By comparing the results from these two estimation methods we obtain an
uncertainty in the spectral type of $\pm$ 1 subtype. Whenever the uncertainty is larger, a colon is added to the spectral subtype.
A ``p" is added to indicate a peculiar spectrum that fits to different subtypes over the wavelength range. This could be the case of non resolved binaries.
An example of the resulting classification for a selection of the best quality spectra (few artifacts and high signal to noise, where possible) for each spectral subtype is plotted in Fig.~\ref{fig.SpectralSequence} to show the spectral sequence.  These spectra are plotted without any smoothing. The missing values are due to artifact removal (as shown in magenta in Fig.~\ref{fig.example}).

\begin{figure}
   \includegraphics[width=0.9\hsize]{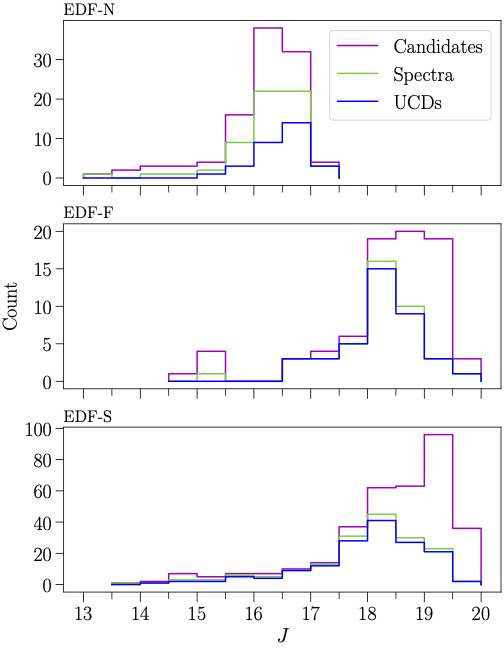} 
   \caption{Histogram of Zhang's compilation divided by field, EDF-N (top), EDF-F (mid), and EDF-S (bottom).  The y-scale is expanded for the latter. Each plot includes all the candidates (magenta line), the dwarfs confirmed by Q1 spectra (green  line) and the UCDs (blue line).  EDF-F and EDF-S cover  fainter candidates than EDF-N (see Sect.~\ref{sc:Jerryspec} for further details).  }
    \label{fig.hist}
\end{figure}

\begin{table}
\caption{\label{tab.xmatch}Number of cross-matched objects between Q1 data and Zhang's compilation).}
\centering
\begin{tabular}{lrrrrr}
\hline\hline
Field & Zhang & Q1 cross-  & Q1 cross- &  Confirmed  &  M7 \\ 
 & candi- & matched & matched  & dwarfs &and \\ 
 & dates &  to $1''$ &  1 - $2.5''$ & & later  \\ 
\hline
EDF-N & 89  & 64  & 22 & 61  &  30  \\ 
EDF-S & 348 & 319 & 28 & 178 & 155 \\  
EDF-F & 78  & 75  & 2  & 42  &  39 \\  
\hline
\end{tabular}
\tablecomments{The fourth column indicates  the number of matches increasing the search radius.  The fifth column gives the total number of dwarfs confirmed. The sixth column gives the number of confirmed dwarfs classified as UCDs. }
\end{table}

\subsection{\label{sc:Jerryspec}Spectra for Zhang's compilation }

We searched for Zhang's compilation of UCD candidates in the Q1 released spectra using their object coordinates. Out of 4.3 million spectra, we found 458 objects matching within 1 arcsecond and 52 objects matching within 1 to 2.5 arcsecond of the coordinates. Together, these 510 objects account for 99\% of Zhang's compilation. An exhaustive comparison of these spectra with standard templates allowed us  to confirm 224 UCDs. Table~\ref{tab.xmatch} shows the number of cross-matched objects, the number of confirmed dwarfs and the number of UCDs among them.  In the case of EDF-N the confirmed UCDs represent 35\% of the candidates with available spectra, given the limited depth of Zhang's compilation in this field    
These numbers represent a minimum, as we  excluded some candidates from confirmation owing to the poor SNR of their spectra. Future visits to the EDFs will allow us to increase the total number of confirmations. Fig.~\ref{fig.hist} shows the distribution of the number of objects per $J$  magnitude. The histograms are divided by field (North, South, and Fornax) to show the objects confirmed spectroscopically in each case. 
Note that from Zhang's compilation,  EDF-F and EDF-S (data from VHS) apparently cover  fainter candidates than EDF-N (data from 2MASS). This is explained by the VHS $J$-band magnitude, which is 3 magnitudes deeper than that from 2MASS $J$-band.    
 The UCDs with the best fits to the standard templates  (negligible fit residuals) are selected as the "UCD benchmarks". There are 60 objects in this list,  37 late-M dwarfs,  21 L dwarfs,   and 2 T dwarfs  with magnitudes
14 $\leq$ $J$  $\leq$ 19. Table~\ref{tab.bench} in Appendix~\ref{sc.bench}  provides the full list.
  The rest of the confirmed objects  include 49 late Ms, 90 Ls, and 6 Ts, most of them with large residuals and 12  that look like UCDs but cannot be reliably classified at the moment.
  The latter group of objects  must wait for future visits to the EDFs which will allow  a better classification, as the survey will gradually gain an additional depth of 2 magnitudes upon completion \citep{EuclidSkyOverview}. 
  
\begin{figure}[h!]
\includegraphics[width=\hsize]{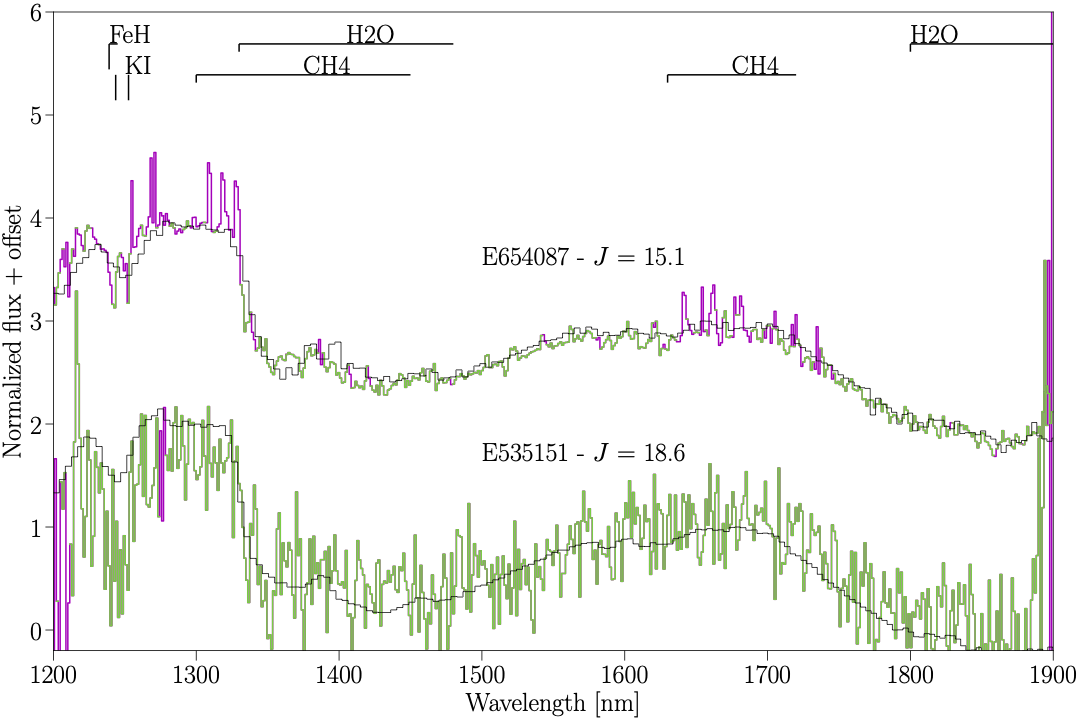} 
   \caption{The brightest and faintest early-L dwarfs  from Zhang's compilation (E654087 and E535151, respectively).  The flux is normalized at 1600~nm. The green lines represent the spectra with \texttt{NDITH}~$>$~2 and  the  magenta lines  indicate the values with  \texttt{QUALITY} $<$~0.7.  The black lines represent the L1 standard template \citep{2004AJ....127.2856B}.  } 
    \label{fig.faint}
\end{figure}

Fig.~\ref{fig.faint} shows the spectra of E535151, the faintest L1  UCD from Zhang's compilation found in Q1 data, together with that of E654087, the brightest   L0  UCD from the same compilation. For this spectral type the span goes from 
15.1 $\leq$ $J$  $\leq$ 18.6.

 \begin{figure}
   \includegraphics[width=1\hsize]{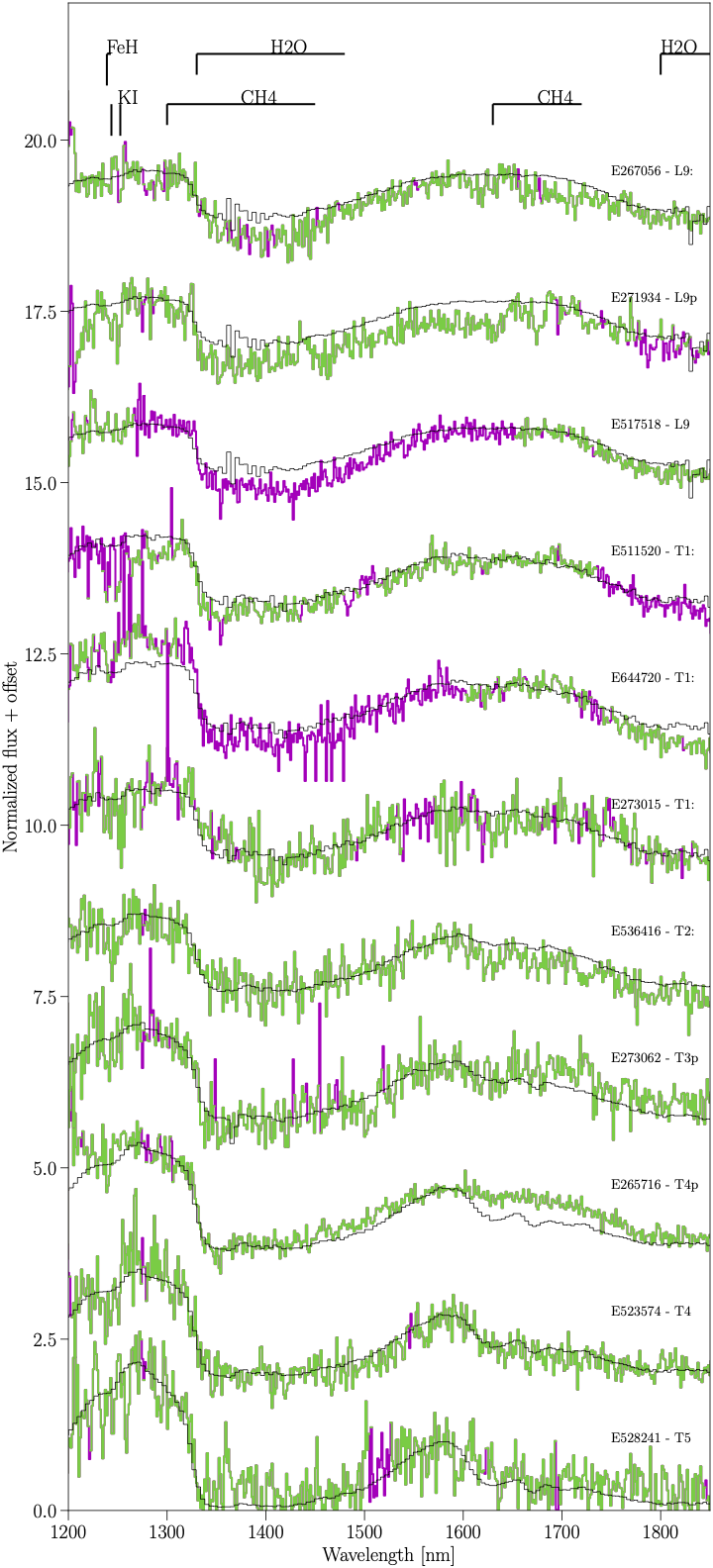}
\caption{Spectra of the new late-L and T dwarfs discovered by the spectral index search.  The flux is normalized at 1600~nm.  The green lines represent the spectra with \texttt{NDITH}~$>$~2 and  the  magenta lines  indicate the values with  \texttt{QUALITY}~$<$~0.7. The black lines are the standard templates (references in Appendix~\ref{sc.splat}).}
\label{fig.ssearch}
\end{figure}

 \begin{figure}
   \includegraphics[width=1\hsize]{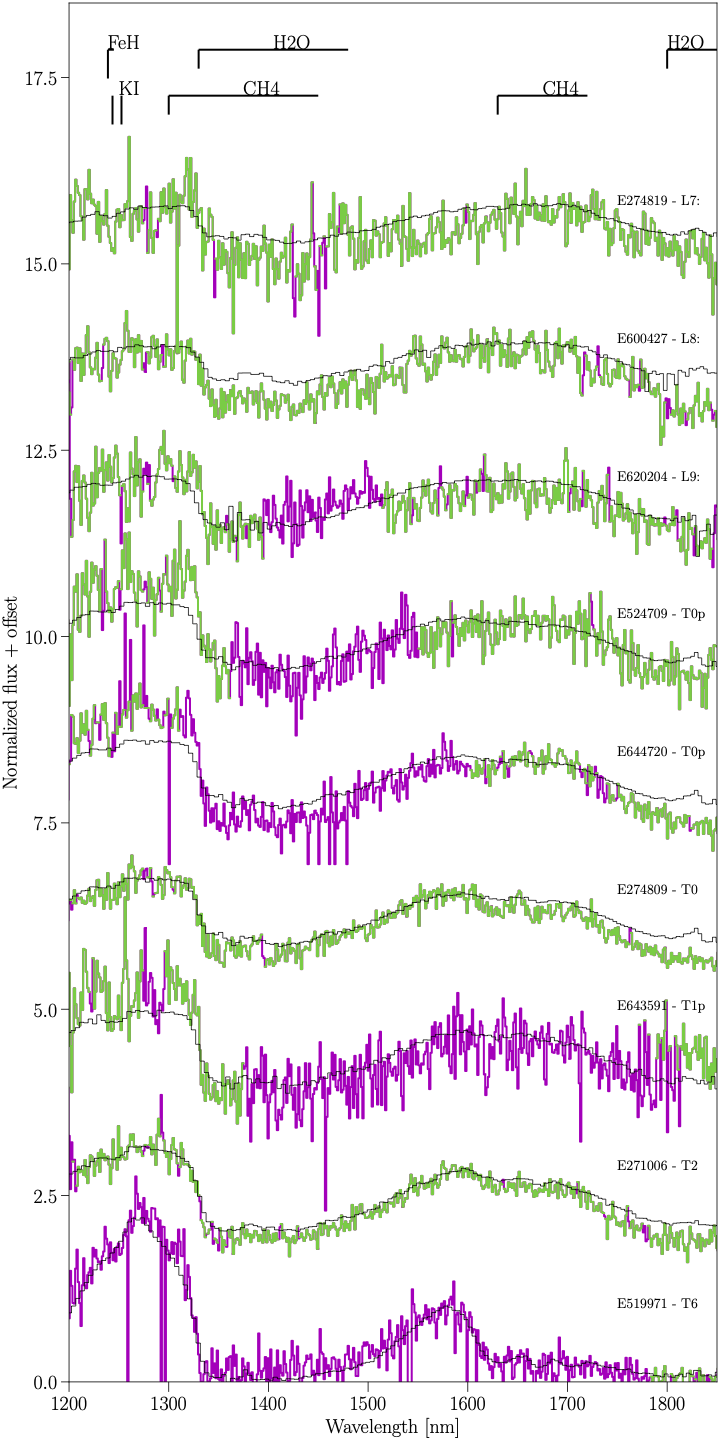}
\caption{Spectra of the new late-L and T dwarfs discovered by {\v Z}erjal's et al. (in prep.) and classified by us (this paper). The flux is normalized at 1600~nm.  The green lines represent the spectra with \texttt{NDITH}~$>$~2 and  the  magenta lines  indicate the values with  \texttt{QUALITY}~$<$~0.7. The black lines are the standard templates (references in Appendix~\ref{sc.splat}).}
\label{fig.smarusa}
\end{figure}

\subsection{\label{sc:resu}Spectra retrieved by the spectral index search   algorithm}

The  11 new late-L and T dwarfs    discovered by the search algorithm were spectroscopically classified and their SpT is included in Table~\ref{tab.directshort}. There  are UCDs for spectral types    from L9 to T5.  Three objects, E265716, E273062, and E271934  present a peculiar spectral energy distribution (signaled with a ``p''),    as shown in
Fig.~\ref{fig.ssearch}. E265716 could be a binary not revealed by the images, owing to its peculiar spectrum that fits a T4 in the $J$-band and T2 in the $H$-band. E271934 is a possible binary as the  \Euclid images show a close companion
 We are investigating this possibility by conducting follow-up observations (Muñoz Torres et al., in prep.).  
There are 5 additional T dwarfs found by the algorithm that were already known: one from Zhang's compilation, two 
found by {\v Z}erjal's et al. (in prep.) in the process of creating their photometric catalog, one from \cite{Mace2013}, and one  found as a QSO contaminant by  Mohandasan et al. (in prep.).
These objects are included in Table~\ref{tab.direct} (Appendix~\ref{sc:listdirect}) with the reference  indicated next to the alias.

\subsection{\label{sc:Marusaspec}Spectra for {\v Z}erjal's catalog  }
We have confirmed and classified  the brighter  L- and T- type UCD candidates from {\v Z}erjal et al. (in prep.), beginning from the   redder ones in the \IE - \YE color. To date, we have classified   more than  one hundred UCDs not included in Zhang's compilation nor in the spectral index search,  of which   4    are T dwarfs,   4   are late-L dwarfs,   56  are early L dwarfs, and   $>$~40  are late M dwarfs.
This list includes some peculiar objects, such as   E643591,    E644720,  and E524709, and   E516938.  
We are investigating their nature. Fig.~\ref{fig.smarusa} shows the spectra of the late L- and T-type UCDs already classified.
 All the new objects confirmed spectroscopically are listed in  Table~\ref{tab.marusa} (Appendix~\ref{sc.marusafull}).

\subsection{\label{sc:notes}Comparing the photometric and  spectral index searches }

\begin{figure}
   \includegraphics[width=1\hsize]{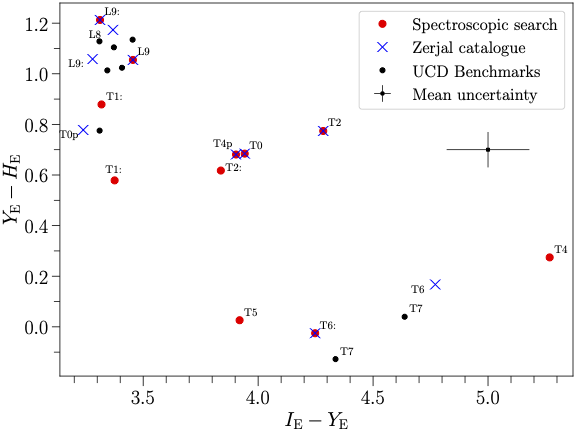} 
   \caption{\YE - \HE vs. \IE - \YE color diagram as in  {\v Z}erjal et al. (in prep.) with the  
 late L- and T-type UCDs. The objects found by the spectral index search  are plotted using the red circles, the UCD benchmarks using the black circles and the ones confirmed from {\v Z}erjal's catalog using blue crosses. The symbol overlap means that the object is found in more than one list.  The color value \IE - \YE = 5.3 for the T4 dwarf has to be taken with caution (see the text).
  }
    \label{fig.color}
\end{figure}

There are   six late-L and T dwarfs   discovered by the spectral index search  that were later   included   {\v Z}erjal's catalog: E265716, E644720, E273015, E517518, E267056, and E271934. This overlap shows the good agreement in the selection criteria. There are also  five T dwarfs  discovered by the spectral index search  algorithm  that are not included in {\v Z}erjal's catalog. Fig.~\ref{fig.color} shows the   \IE - \YE and   \YE - \HE colors of the UCDs found by the two searches.
The objects found exclusively in the spectral index search  are:

\begin{description}
\item[E528241] This object, classified as a T5,  is very faint in the visible ($ I_{E}$ = 27.6) and has  colors  \IE - \YE = 4 and   \YE - \HE = 0. Even if its colors match the {\v Z}erjal et al. (in prep.) criteria it was not included in the photometric selection owing to  the \texttt{SNR\_VIS} flag~\footnote{\url{https://euclid.esac.esa.int/dr/q1/dpdd/}}.

\item[E523574] This object is classified as a T4. It is   faint in the visible ($ I_{E}$ = 26.7), with a color    \YE - \HE = 0.3.  It was not included in the photometric selection because of the \texttt{vis\_det}  flag in VIS data. Thus, its color  \IE - \YE = 5.3 has to be taken with caution.  The spectrum of this object is also included in Fig.~\ref{fig.SpectralSequence}.

\item[E273062] This object is classified as a T3p owing to its peculiar spectrum, probably due to a close very bright object.  The flags (\texttt{det\_quality\_flag, flag\_vis, flag\_y}, and \texttt{flag\_h}) for the images are activated, so the color values have to be treated with caution. 

\item[E536416] This object, classified as a T2:,  is  faint in the visible ($ I_{E}$ = 25.4), with  colors  \IE - \YE = 3.8 and   \YE - \HE = 0.6. It was not included in the photometric selection because of the ellipticity filtering flag.

\item[E644720] This object is classified as a T1: and  is relatively bright in the visible ($ I_{E}$ = 23.2), with  colors  \IE - \YE = 3.0 and   \YE - \HE = 0.5. It was filtered out from the photometric selection because of the \texttt{mumax\_minus\_mag} value.

\end{description}

From these objects we note that the main limitation is the quality of the images and derived data, given that the colors and filtering parameters are based on these data.

Two objects  included in both the spectral index search and the photometric catalog are worth mentioning. 
 E265716 is a possible unresolved binary owing to its peculiar spectrum that fits a T4 in the $J$-band and T2 in the $H$-band.   Its color values are:   \IE - \YE = 3.9 and   \YE - \HE = 0.7.   The second object,  E271934 is a possible binary with a close companion that appears to be resolved in  \Euclid images.  Its peculiar spectrum fits an L9 in the $J$-band and L4 in the $H$-band.  Its color values are:   \IE - \YE = 3.5 and   \YE - \HE = 0.8.
 A full discussion is included in Muñoz Torres et al. (in prep.).

There are four objects selected by the photometric search that were not identified by the spectral index search:  

\begin{description}
\item[E600427] This object is an L8: dwarf that has P$J$-$JH$  and  P$H$-$JH$ values    
below the minimum ($>$0.6 and  $>$0.5, respectively).

\item[E620204] This object is  an L9:  dwarf that also has  P$J$-$JH$ and P$H$-$JH$  values  
below the minimum, owing to  the presence of artifacts or contamination around 1400~nm.

\item[E519971] This object  is a T5:  dwarf that was filtered out because of the presence of artifacts at different wavelengths producing NaN values in the search results.

\item[E524709] This object  is a peculiar T0 dwarf that was filtered out owing to the presence of artifacts or contamination around 1400~nm.

\end{description}

\begin{figure*}
   \centering
 \includegraphics[width=0.7\linewidth]{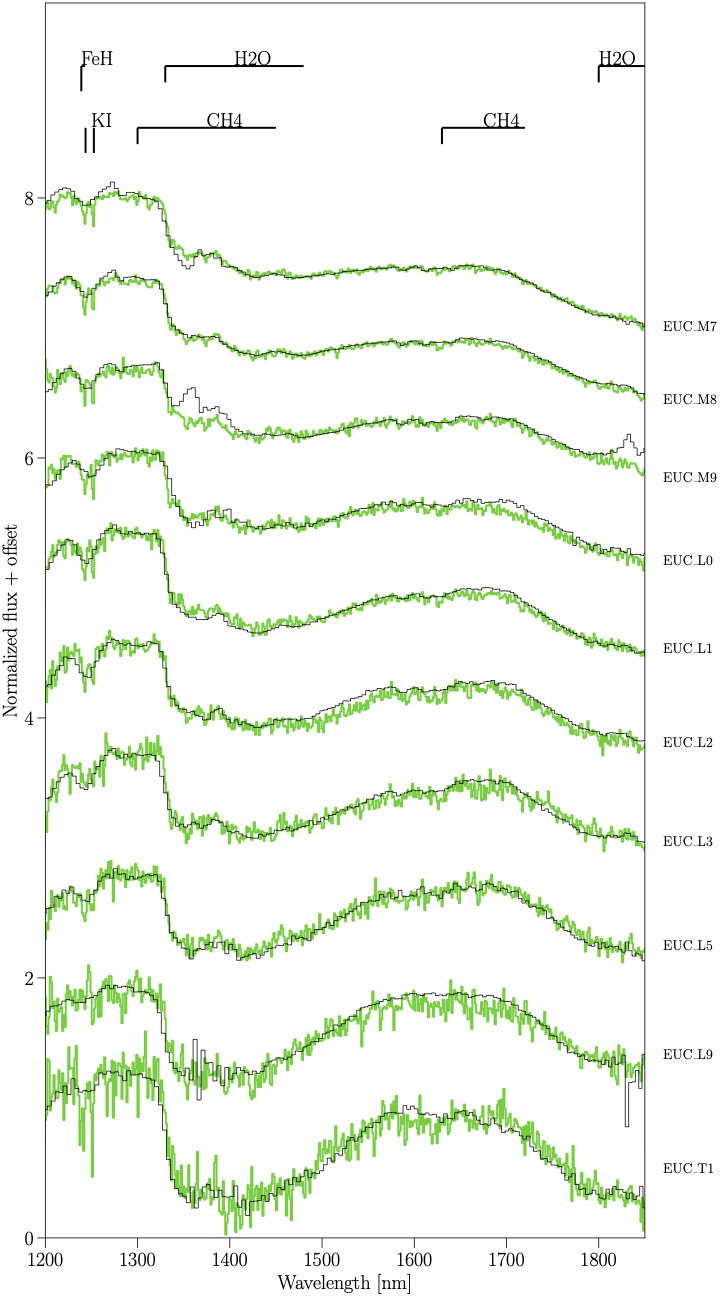}  
\caption{Preliminary templates (green lines) obtained from the combination by median of multiple \Euclid spectra. The flux is normalized at 1600~nm.   Each spectrum is named using "EUC" plus its spectral type.  The black lines are the standard templates of the same spectral type (references in Appendix~\ref{sc.splat}). }
\label{fig.templates}
\end{figure*}

\section{\label{sc:template}Preliminary \textit{EUCLID}\, UCD templates  }
Spectra obtained with \Euclid have the advantage of being free from telluric absorption bands. The large number of objects of every spectral type that we are finding and classifying provides an opportunity to create new UCD templates observed with \Euclid from space. 
From the available spectra, we  combined the best spectra of every spectral subtype  to create such templates.
Fig.~\ref{fig.templates} shows the first step in this direction, 
a comparison  of the \Euclid combined spectra with  ground-based standard templates (black lines; references in Appendix~\ref{sc.splat}).  Note the differences in the regions of the telluric absorption bands. These templates are the result of the combination by median of more than 10 spectra for the spectral types M7, M8, L0, and L1. The rest of templates have a smaller number (see Table~\ref{tab.tefftempl}), thus the result is  noisier. Future data releases will allow us to improve these templates by combining a larger number of spectra.

\begin{table}
\caption{\label{tab.tefftempl}Preliminary templates.}
\centering
\begin{tabular}{lrr}
\hline\hline
Alias &   Number of  & $T_{\mathrm{eff}}$\\
     & spectra & (K) \\
\hline
\noalign{\smallskip}
EUC\_T1   & 3  & $1536_{-59}^{+93 \phantom{0}}$  \\
\noalign{\smallskip}
EUC\_L9  &  4   &  $1453_{-129}^{+61}$  \\
\noalign{\smallskip}
EUC\_L5   & 3  &  $1504_{-139}^{+71}$  \\
\noalign{\smallskip}
EUC\_L3  &  6   &  $1705_{-118}^{+54}$  \\
\noalign{\smallskip}
EUC\_L2  &  8  &  $1799_{-51}^{+61 \phantom{0}}$  \\
\noalign{\smallskip}
EUC\_L1  &  14   &  $2031_{-32}^{+140}$  \\
\noalign{\smallskip}
EUC\_L0  &  11   &  $2264_{-37}^{+108}$  \\
\noalign{\smallskip}
EUC\_M9  &  8   &  $2264_{-51}^{+123}$  \\
\noalign{\smallskip}
EUC\_M8  &  29   &  $2376_{-35}^{+96 \phantom{0}}$  \\
\noalign{\smallskip}
EUC\_M7    & 13  & .  .  . \phantom{40} \\

\hline
\end{tabular}
\tablecomments{The second column indicates the number of spectra combined to obtain the template. The  method to estimate  the effective temperature ($T_{\mathrm{eff}}$) is described in Sect.~\ref{sc:teff} }
\end{table}

\section{\label{sc:anal} Spectral analysis }

\subsection{Effective temperature determination} \label{sc:teff}

We determined the effective temperature ($T_{\mathrm{eff}}$) for objects with a spectral type equal or later than M8. For this, we adopted the same deep transfer learning approach presented by \citet{MASBUITRAGO2024}, using a combination of autoencoder and convolutional neural network architectures, and tailored it to the low-resolution domain. The autoencoder models are trained on a synthetic grid, based on the \texttt{Sonora Elf Owl} models \citep{elfowl}, and then fine-tuned using our sample of \Euclid UCD spectra. An in-depth discussion of the adaption of our methodology to the low-resolution domain is presented by Mas-Buitrago et al. (in prep.).

\begin{figure}
   \centering
   \includegraphics[width=0.9\hsize]{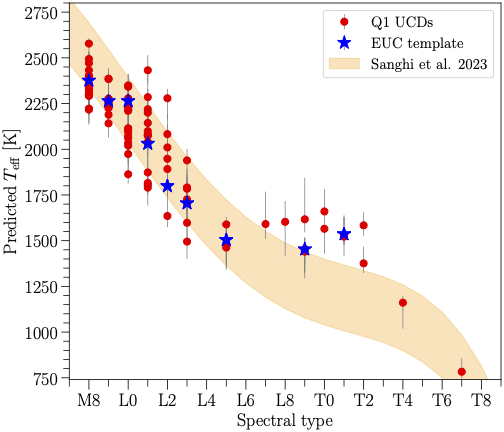} 
   \caption{Predicted effective temperatures as a function of spectral classification for individual UCDs (red circles) and \Euclid UCD templates (blue stars; see Sect.~\ref{sc:template}). The shaded orange area indicates the semi-empirical relation and RMS scatter derived by \citet[][see Fig. 16]{Sanghi2023} for field UCDs.}
    \label{fig.teff}
\end{figure}

Figure \ref{fig.teff} shows the relationship of the derived $T_{\mathrm{eff}}$ values with the spectral classification from Section \ref{sc:class}. The full list of measurements can be found in 
Table~\ref{tab.teffalll}~(Appendix~\ref{sc.appTeff}).
We also include the $T_{\mathrm{eff}}$ of the  available \Euclid UCD templates (which are not affected by artifacts, see  Table~\ref{tab.tefftempl} and Sect.~\ref{sc:template}).
The values for the  UCDs  earlier than L7 are consistent  with the semi-empirical measurements by \cite{Sanghi2023}, also shown in the Figure.
For the L-T transition, the predicted effective temperatures seem to be higher than those from \cite{Sanghi2023}, showing a different profile up to T2.  Beyond that, the agreement is again very good. Future observations and data releases will increase the number of objects and,  with the incorporation of  the missing spectral subtypes, we will be able to confirm this behavior.

\subsection{\label{sc:indice}Spectral indices}

\begin{figure}
\includegraphics[width=\hsize]{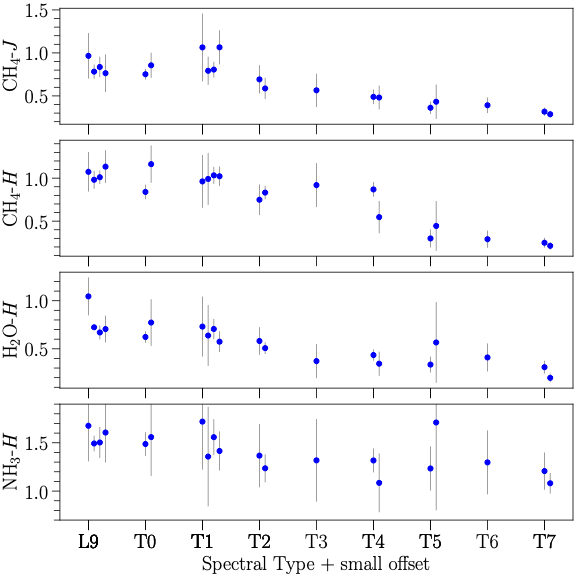} 
   \caption{Spectral indices for late L- and T-type UCDs. The three top panels show the CH$_4$-$J$, CH$_4$-$H$, and  H$_2$O-$H$  NIR spectral indices defined by \cite{Burgasser06}. The bottom panel shows the NH$_3$-$H$ spectral index as defined by \cite{EGE2021amonia}.}
    \label{fig.nh3}
\end{figure}

We measured the CH$_4$-$J$, CH$_4$-$H$, and  H$_2$O-$H$  NIR spectral indices defined by \cite{Burgasser06} for UCDs with spectral types later than L8. Fig.~\ref{fig.nh3} shows how the values of these indices decrease with respect to the spectral type, in good agreement with the behavior found by \cite{Burgasser06}, and other works \citep[see e.g.][]{ emily2017, Lodieu2018}.

We also measured the NH$_3$-$H$ spectral index that was redefined by \cite{EGE2021amonia} anticipating \Euclid low spectral resolution. This index is an indicator of the absorption of ammonia in late T and Y-type UCDs. We include earlier subtypes to investigate the dispersion of values obtained with   \Euclid spectra.

\begin{figure}
   \centering
   \includegraphics[width=0.8\hsize]{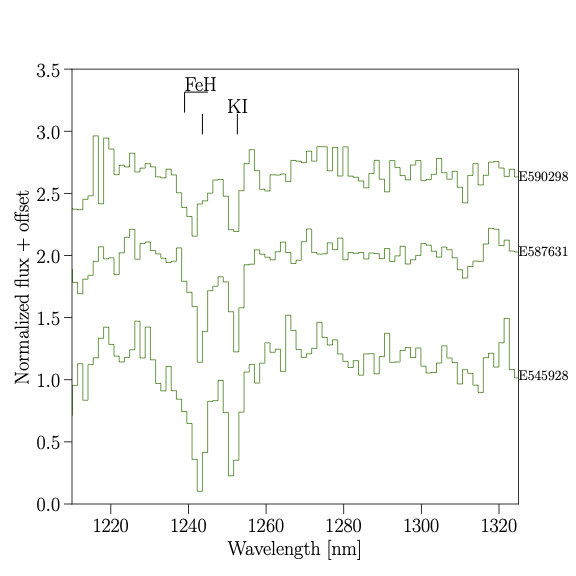} 
   \caption{Example of different strengths in the K{\small{I}}~absorption doublet at 1243.6 /  1252.6~nm (air wavelength) for
   three M8 dwarfs from Q1. }
    \label{fig.K1ewzoom}
\end{figure}

\begin{figure}
\includegraphics[width=\hsize]{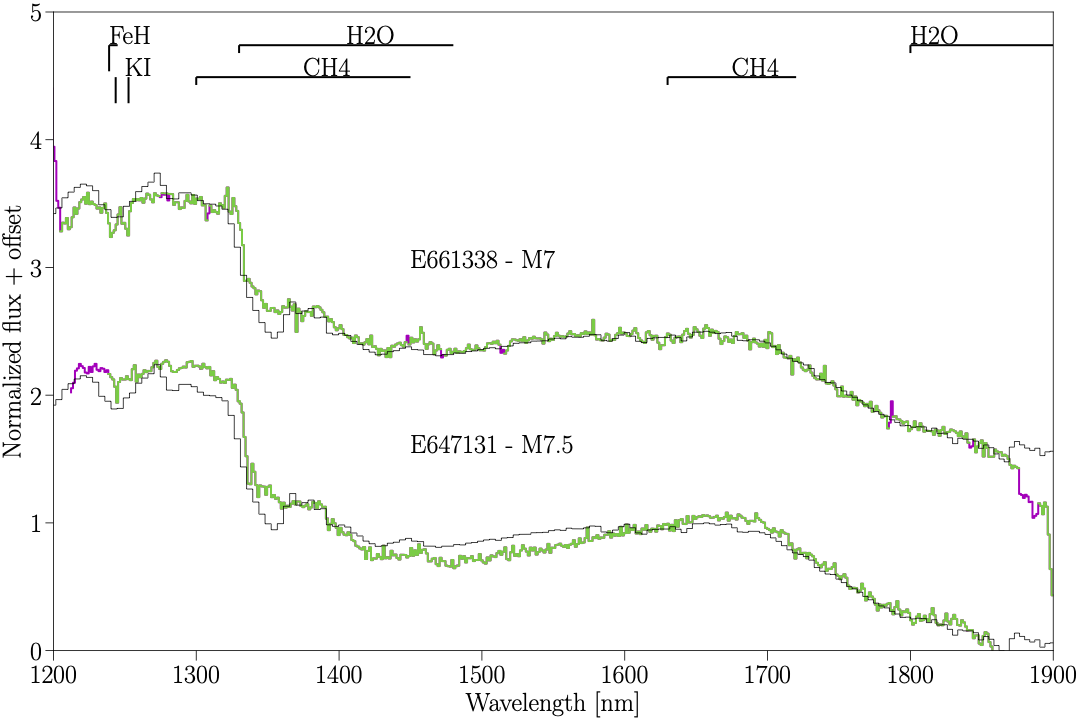} 
   \caption{Comparison of old- and young-object spectra, both obtained with \Euclid. Note the absence of K{\small{I}}~absorption doublet in the lower spectrum, which corresponds to a young population. The flux is normalized at 1600~nm.  E661338  spectrum is included in Q1 data of EDF-S, while that of E647131, a young object from  ERO-02 regions, is taken from Dominguez-Tagle et al. (in prep.). The black lines represent the M7 standard template \citep{burgasser2008}. }
    \label{fig.age}
\end{figure}

\subsection{\label{sc:age}K{\small{I}}~absorption doublet}
NISP spectral range includes the region of the  K{\small{I}}~absorption doublet at 1243.6 /  1252.6~nm (air wavelength), and the spectral resolution is good enough to resolve the two lines.  This potassium doublet is a known spectral feature that increases its depth with high gravity and therefore is related to age \citep{speclassbook}. Fig.~\ref{fig.K1ewzoom} shows an example of three M8 dwarfs from Q1 with different strengths in the potassium doublet. 
Given the different properties of the stellar populations in the EDFs and in the ERO-02 fields  \citep{ege25},  we can compare old and young UCD populations. Fig.~\ref{fig.age} shows  two M7 objects as an example,  E661338 from the EDF-S    and  E647131, a young object from the ERO-02 regions. The latter spectrum is taken from Dominguez-Tagle et al. (in prep.). 
The potassium doublet is strong in  object E270809 whereas it is not present in E647131, a UCD from a young population.
E647131 spectrum was  extracted using our own code, also described in Dominguez-Tagle et al. (in prep.).

\begin{figure}
   \centering
   \includegraphics[width=1\hsize]{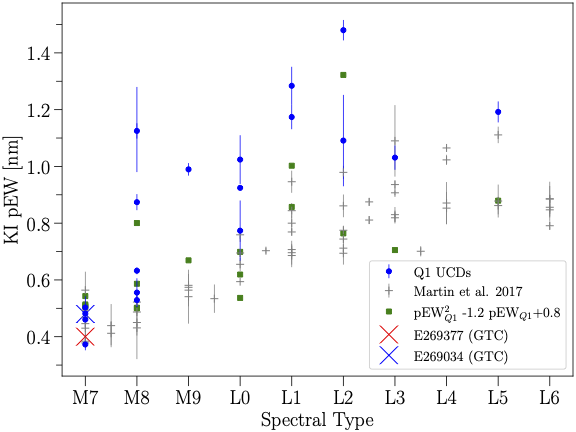} 
   \caption{K{\small{I}}~1252.6~nm pseudo EWs as a function of the spectral type   
     for the objects listed in Table~\ref{tab.rvel} (blue circles). 
   The green boxes represent the 
   expected values for higher resolution spectra,
  following \cite{Malkan2002} procedure (the legend shows the empirical relation).   
   The blue and red $\times$ symbols indicate the pEW for E269377 from Q1 and GTC \citep{ZANG2024}, respectively. 
   The grey crosses represent the field UCDs from  \cite{emily2017}.  }
    \label{fig.K1ew}
\end{figure}

We  measured the pseudo equivalent width (pEW) of the K{\small{I}}~absorption line  1252.6~nm in the spectra of 20 dwarfs that present the K{\small{I}}~absorption doublet region not affected by artifacts. This list includes dwarfs from M7 to L4, and the results are presented in Table~\ref{tab.rvel}.
To measure the pEW we deblend the two potassium lines and use a window of 20~nm for the pseudo continuum   out of the line region.  We do not estimate the pEW for the line at 1243.6~nm because it could be blended with an FeH feature at 1239~nm, which increases through the late-M dwarfs to early-L dwarfs \citep{allers13}. Fig.~\ref{fig.K1ew} shows the pEW values as a function of the spectral type. 

Object E269377 (WISEA J175730.71+674138.4) was observed with GTC/EMIR ($R\sim 1000$)  by  \citet[][see Sect.~\ref{sc:dataJerry}]{ZANG2024}, and the values from both spectra measured in the same manner are included in the figure using $\times$ symbols. 
 The $\sim$0.1~nm difference between both measurements is likely due to the different spectral resolutions,  as reported by  \cite{Malkan2002}. Three objects show higher pEWs:  E545928 (M8), E529572 (M8), and E619358 (L2). 
E529572 and E619358 present strong K{\small{I}}~1252.6~nm lines which could indicate a slightly higher gravity or a difference in the spectral subtype. Additional observations are needed to confirm this possibility.

Fig.~\ref{fig.K1ew} includes the measurements for the field UCDs reported by  \cite{emily2017} plotted as gray dots. Other authors show similar results \citep[see e.g.][]{allers13, Lodieu2018, Burgasser2025}.  
In contrast, our pEW values are systematically higher, although they present a similar scatter and follow the same correlation with respect to the spectral type. In order to verify any systematic effect, we took the $R\sim 2000$ spectra observed by \cite{McLean2003} of the objects reported by  \cite{emily2017} and measured the K{\small{I}}~1252.6~nm  pEW using the same method applied to Q1 data.
The differences among the pEW values obtained with our method and those by \cite{emily2017} were small, with a mean value of 0.06~nm. Then, following \cite{Malkan2002} procedure, we measured again the pEW after degrading the spectral resolution to match that of NISP ($R\sim 450$). These new values were systematically higher relative to the original ones. We computed a second-order polynomial fit between the values of the original and the degraded spectra. The polynomial coefficients are shown in the legend of the figure. We applied this relation to our measurements of Q1 spectra to derive the expected values for a similar resolution to that of  \cite{emily2017}. The agreement between both sets of measurements is reasonable. The most plausible explanation is the blending of the two potassium lines (1243.6  and  1252.6~nm) owing to the low NISP spectral resolution.  This effect needs to be considered in future pEW measurements from \Euclid spectra.

\begin{table}[h!]
\caption{\label{tab.rvel} K{\small{I}} 1252.6~nm  measurements for selected objects }
\centering
\begin{tabular}{lllrc}
\hline\hline
Alias &  SpT & K{\small{I}} pEW  & $K{\small{I}} \ (\lambda - \lambda_0 ) $           & Field      \\ 
      &      &  (nm)   &   (nm) &   \\ 
\hline
  E576081 & L5:  & 1.19 $\pm$  0.04 & 0.58 $\pm$  0.08 & S\\
  E270698 & L3:  & 1.09 $\pm$  0.16 & -0.60 $\pm$  0.08 & N\\
  E574238 & L2:  & 1.03 $\pm$  0.04 & -0.01 $\pm$  0.14 & S\\
  E619358 & L2:  & 1.48 $\pm$  0.04 & -0.60 $\pm$  0.07 & S\\
  E609567 & L1 & 1.28 $\pm$  0.07 & -0.76 $\pm$  0.21 & S\\
  E274793 & L1   & 1.17 $\pm$  0.04 & -0.10 $\pm$  0.07 & N\\
  E524930 & L0 & 1.02 $\pm$  0.09 & 0.55 $\pm$  0.25 & F\\
  E265832 & L0 & 0.92 $\pm$  0.01 & -0.14 $\pm$  0.03 & N\\
  E266937 & L0:   & 0.77 $\pm$  0.11 & -0.55 $\pm$  0.38 & N\\
  E638486 & M9 & 0.99 $\pm$  0.02 & -0.42 $\pm$  0.06 & S\\
  E529572 & M8 & 0.87 $\pm$  0.03 & -0.83 $\pm$  0.07 & F\\
  E545928 & M8 & 1.13 $\pm$  0.15   & -0.20 $\pm$  0.07 & F\\
  E658543 & M8:  & 0.53 $\pm$  0.02 & -0.15 $\pm$  0.04 & S\\
  E590298 & M8:  & 0.56 $\pm$  0.05 & -0.34 $\pm$  0.08 & S\\
  E587631 & M8 & 0.63 $\pm$  0.01 & 0.26 $\pm$  0.10 & S\\
  E273794 & M7 & 0.37 $\pm$  0.02 & -0.06 $\pm$  0.23 & N\\
  E269034 & M7 & 0.50 $\pm$  0.04 & -1.20 $\pm$  0.09 & N\\
  E269034$^{GTC}$  & M7 & 0.31 $\pm$  0.01 & -0.12 $\pm$  0.02 & N  \\
  E269377 & M7 & 0.48 $\pm$  0.02 & 0.07 $\pm$  0.14 & N\\
   E269377$^{GTC}$ & M7 & 0.38 $\pm$  0.04 & -0.45 $\pm$  0.05 & N\\
  E661338 & M7 & 0.46 $\pm$  0.03 & -0.82 $\pm$  0.19 & S\\
\hline
\end{tabular}
\tablecomments{
  The measurements for objects observed with GTC/EMIR are marked with "GTC".
Uncertainties in the spectral classification $> 1$ subtype are noted by a colon. 
The fifth column N, F, and S letters reefer to EDF-N, EDF-F, and EDF-S, respectively. 
  }
\end{table}

\begin{figure}
   \centering
   \includegraphics[width=1\hsize]{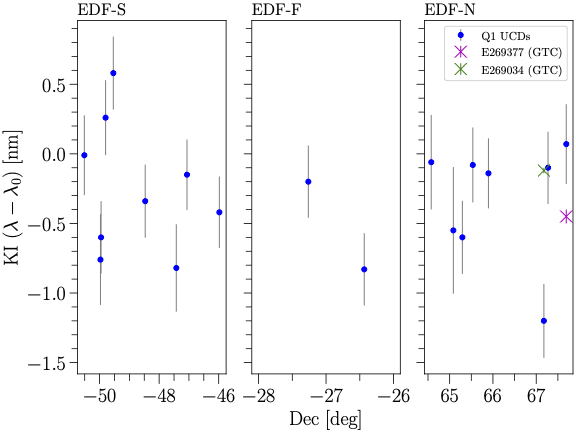} 
   \caption{K{\small{I}}~1252.6~nm observed wavelength shift for the objects listed in Table~\ref{tab.rvel}. They are plotted as a function of  their declination to group them per field.  Left, mid, and right panels correspond the EDF-S, EDF-F, and EDF-N, respectively.  
   The  magenta and green    $\times$ symbols show the wavelength shift for two objects observed also with GTC/EMIR, E269377 by \cite{ZANG2024} and E269034 by Muñoz Torres et al. (in prep.), respectively.   }
    \label{fig.K1wave}
\end{figure}

The K{\small{I}}~1252.6~nm absorption line could also allow to estimate  radial velocities, within the limits of the NISP low spectral resolution and calibration errors. The measurements of the wavelength accuracy reported by \cite{Q1-TP006} indicate that $>$~80\% of the occurrences are below 0.25~nm. From this value, the minimum 3$\sigma$ reliable velocity estimate is 180~$\rm km~s^{-1}$ .
Therefore, we do not aim to give an absolute value for the radial velocity, but to  identify any object whose radial velocity deviates significantly from that of the rest of objects by comparing their $K{\small{I}} \ (\lambda - \lambda_0 ) $ values.   Table~\ref{tab.rvel} and Fig.~\ref{fig.K1wave} show the results. The figure is split  into three panels, one per field. 
The objects in EDF-N and EDF-S present a  dispersion  of 0.4 and 0.5~nm RMS, respectively, which is larger than the typical wavelength accuracy reported by \cite{Q1-TP006}. The maximum difference between two objects in the EDF-S is 1.4~nm, that would be more than 330~$\rm km~s^{-1}$. 
E269034, located in EDF-N, also appears to deviate from the main group.  However, a recent follow-up observation with GTC/EMIR shows only a very small shift (green  $\times$ symbol).
The large difference between the two measurements exceeds the minimum 3$\sigma$ threshold for a reliable value, based on the available wavelength accuracy report.  Therefore, the current measurements from NISP spectra should be interpreted with caution. 
Further observations, either from ground-base telescopes or future visits of \Euclid to these fields are needed to confirm these measurements.  In the latter case, they will gradually improve the SNR, the wavelength calibration, and hence, the K{\small{I}}  measurements.

\section{\label{sc:concl}Conclusions  }

We searched Q1 data for the UCD candidates from Zhang's compilation and confirmed   the UCD nature of   224 objects. We  selected 60  UCD benchmarks from this list to be the reference for the photometric selection criteria. The list includes  37 late M dwarfs,  21 L dwarfs, and 2 T dwarfs. 

We  also  searched for late-L and T-type UCDs directly in Q1 spectra and discovered  11  objects (8 T dwarfs and 3 late-L dwarfs). 

Out of 5300 candidates in {\v Z}erjal's catalog,  we have (to date) confirmed  more than  one hundred UCDs that were not included in Zhang's compilation, nor in the spectral index search.   We compared the photometric and spectral index searches  offering hints as to why some of them  were not identified by the photometric selection.
 
In total, we  discovered and classified  12  T dwarfs and  7  late-L dwarfs, combining the photometric and spectral index searches.  

We created a preliminary list of \Euclid UCD templates from M7 to  T1 , that can be used as a reference for the future search and classification of candidates.
 
We  demonstrated that NISP's sensitivity is sufficient to get a useful spectrum of UCDs down to magnitude $J \sim 19 $   from a single visit to a field or ROS. We  investigated the capabilities of \Euclid to study the physical properties of UCDs by their spectral features, specifically H$_2$O, CH$_4 $, and NH$_3$ indices and the pEW of the reddest potassium line in the $J$~band.  

This paper is only a first step in the study of \Euclid UCDs  and will be improved with each subsequent data release in terms of number of objects and analysis quality, as the survey will gain  2 magnitudes in depth upon completion \citep{EuclidSkyOverview}.

\begin{acknowledgments}
We are grateful to T. Dupuy for his suggestions in an early version of the paper. We thank P. Schneider for the style suggestions and T. Mahoney for the English grammar corrections.
Funding for CDT, M{\v Z},  NS, ELM, NV, SM, and TS was provided by the European Union (ERC Advanced Grant, SUBSTELLAR, project number 101054354). DB, VB, PC, NL, EM, PMB, ES and JZ acknowledge financial support from the Agencia Estatal de Investigaci\'on (AEI/10.13039/501100011033) of the Ministerio de Ciencia e Innovaci\'on  through the following projects:
 PID2022-137241NB-C41 (for VB, NL, EM, and JZ), 
 PID2020-112949GB-I00 (for PMB, PC and ES;  Spanish Virtual Observatory https://svo.cab.inta-csic.es), and, 
 PID2023-150468NB-I00 (for DB).
The authors wish to acknowledge the contribution of the IAC High-Performance Computing support team and hardware facilities to the results of this research. PMB acknowledges support from the Instituto Nacional de
Técnica Aeroespacial through grant PRE-OVE. 
NPB is funded by Vietnam National Foundation for Science
and Technology Development (NAFOSTED) under grant number 103.99-2020.63.
This work has made use of the \Euclid Quick Release
  Q1 data from the \Euclid\/ mission of the European Space Agency
  (ESA),  \cite{https://doi.org/10.57780/esa-2853f3b}. 
This work has made use of the Early Release
  Observations (ERO) data from the \Euclid\/ mission of the European
  Space Agency (ESA), 
  \cite{https://doi.org/10.57780/esa-qmocze3}. 
The authors acknowledge the Euclid Consortium,
  the European Space Agency, and a number of agencies and
  institutes that have supported the development of \Euclid, in
  particular
  the Agenzia Spaziale Italiana,
  the Austrian Forschungsf\"orderungsgesellschaft funded through BMK,
  the Belgian Science Policy,
  the Canadian Euclid Consortium,
  the Deutsches Zentrum f\"ur Luft- und Raumfahrt,
  the DTU Space and the Niels Bohr Institute in Denmark,
  the French Centre National d'Etudes Spatiales,
  the Funda\c{c}\~{a}o para a Ci\^{e}ncia e a Tecnologia,
  the Hungarian Academy of Sciences,
  the Ministerio de Ciencia, Innovaci\'{o}n y Universidades,
  the National Aeronautics and Space Administration,
  the National Astronomical Observatory of Japan,
  the Netherlandse Onderzoekschool Voor Astronomie,
  the Norwegian Space Agency,
  the Research Council of Finland,
  the Romanian Space Agency,
  the State Secretariat for Education, Research, and Innovation (SERI) at the Swiss
  Space Office (SSO),
  and the United Kingdom Space Agency.
  A complete and detailed list is available on the \Euclid\ web site
  (\url{www.euclid-ec.org}).

This work is partly based on the observing proposal GTCMULTIPLE2D-25A, made with the Gran Telescopio Canarias (GTC), installed at the Spanish Observatorio del Roque de los Muchachos of the Instituto de Astrofísica de Canarias, on the island of La Palma. This work is partly based on data obtained with the instrument EMIR, built by a Consortium led by the Instituto de Astrofísica de Canarias. EMIR was funded by GRANTECAN and the National Plan of Astronomy and Astrophysics of the Spanish Government.
We thank the anonymous referee for their insightful comments which improved the paper.

\end{acknowledgments}





%
\facilities{\Euclid space mission, GTC, VLT}

\software{SPLAT \citep{SPLAT}, astropy \citep{2013A&A...558A..33A,2018AJ....156..123A,2022ApJ...935..167A}, TOPCAT \citep{2005ASPC..347...29T}, splinecoeff (\url{github.com/nikolavitas/idl/}). 
          }


\appendix

\section{\label{sc.splat} UCD standard templates}       

\begin{table}[h!]
\caption{\label{tab.std} Standard templates used in this paper}
\centering
\begin{tabular}{llll}
\hline\hline
SpT  &  Name  &    SpT reference  &  Data reference \\

\hline
M6.0  &  Wolf 359  &    \cite{Kirkpatrick1991}  &  \cite{burgasser2008} \\
M7.0  &  VB 8  &    \cite{Kirkpatrick1991}  &  \cite{burgasser2008} \\
M8.0  &  VB 10  &    \cite{Kirkpatrick1991}  &  \cite{2004AJ....127.2856B} \\
M9.0  &  LHS 2924  &    \cite{Kirkpatrick1991}  &  \cite{2006AJ....131.1007B} \\
L0.0  &  2MASP J0345432+254023  &    \cite{Kirkpatrick99}  &  \cite{2006AJ....131.1007B} \\
L1.0  &  2MASSW J1439284+192915  &   \cite{Kirkpatrick99}  &  \cite{2004AJ....127.2856B} \\
L2.0  &  Kelu-1  &    \cite{Kirkpatrick99}  &  \cite{2007ApJ...658..557B} \\
L3.0  &  2MASSW J1506544+132106  &    \cite{2000AJ....120.1085G}  &  \cite{2007ApJ...659..655B} \\
L4.0  &  2MASS J21580457-1550098  &  \cite{2008ApJ...689.1295K}  &  \cite{2014ApJ...794..143B} \\
L5.0  &  SDSS J083506.16+195304.4  &    \cite{2006AJ....131.2722C}  &  \cite{2006AJ....131.2722C} \\
L6.0  &  2MASSI J1010148-040649  &   \cite{2003AJ....126.2421C}  &  \cite{2006AJ....132..891R} \\
L7.0  &  2MASSI J0103320+193536  &    \cite{2022AandA...664A.111B}  &  \cite{2003AJ....126.2421C} \\
L8.0  &  2MASSW J1632291+190441  &    \cite{Kirkpatrick99}  &  \cite{2004AJ....127.2856B} \\
L9.0  &  DENIS-P J0255-4700  &    \cite{2008ApJ...689.1295K}  &  \cite{Burgasser06} \\
T0.0  &  SDSS J120747.17+024424.8  &    \cite{2002AJ....123.3409H}  &  \cite{2007AJ....134.1162L} \\
T1.0  &  SDSSp J083717.22-000018.3  &    \cite{Burgasser06}  &  \cite{Burgasser06} \\
T2.0  &  SDSSp J125453.90-012247.4  &   \cite{Burgasser06}  &  \cite{2004AJ....127.2856B} \\
T3.0  &  2MASS J12095613-1004008  &    \cite{Burgasser06}  &  \cite{2004AJ....127.2856B} \\
T4.0  &  2MASSI J2254188+312349  &   \cite{Burgasser06}  &  \cite{2004AJ....127.2856B} \\
T5.0  &  2MASS J15031961+2525196  &   \cite{Burgasser06}  &  \cite{2004AJ....127.2856B} \\
T6.0  &  SDSSp J162414.37+002915.6  &   \cite{Burgasser06}  &  \cite{2004AJ....127.2856B} \\
T7.0  &  2MASSI J0727182+171001  &    \cite{Burgasser06}  &  \cite{2004AJ....127.2856B} \\
T8.0  &  2MASSI J0415195-093506  &    \cite{Burgasser06}  &  \cite{2004AJ....127.2856B} \\
T9.0  &  UGPS J072227.51-054031.2  &   \cite{Burgasser06}  & \cite{SPLAT} \\
\hline
\end{tabular}
\end{table}

\newpage

\section{\label{sc.bench}UCD benchmarks}
Description in Sect.~\ref{sc:Jerryspec}

\begin{table}[h!]
\digitalasset
\caption{\label{tab.bench} Descriptive version of the ``Full list of UCD benchmarks" table.  }
\centering
\begin{tabular}{llrrrc}
\hline\hline
Alias    &    SpT       &    \Euclid  ID   &    RA(J2000)   &    Dec(J2000)    &    $J$ \\ 
    &           &      &       &      &     (mag) \\ 
\hline
  E529572 & M8 & -529572577269202972 & 03:31:49.7 & -26:55:13.1 & 16.8\\
  E531137 & M8 & -531137796271166336 & 03:32:27.3 & -27:06:59.9 & 18.0\\
  E531382 & L1 & -531382284267016817 & 03:32:33.2 & -26:42:06.1 & 17.5\\
  E535151 & L1: & -535151413283919886 & 03:34:03.6 & -28:23:31.2 & 18.6\\
  E545928 & M8 & -545928778272623795 & 03:38:22.3 & -27:15:44.6 & 17.2\\
  E564140 & L2: & -564140829495758229 & 03:45:39.4 & -49:34:33.0 & 18.0\\
  E566608 & L1: & -566608621486213957 & 03:46:38.6 & -48:37:17.0 & 17.6\\
  E573377 & M8 & -573377256498994280 & 03:49:21.1 & -49:53:57.9 & 17.0\\
  \hline
\end{tabular}
\tablecomments{
 Uncertainties in the spectral classification $> 1$ subtype are noted by a colon.
$J$ magnitudes are from Zhang's compilation. 
This table is published in its entirety in the electronic 
edition of the {\it Astrophysical Journal}.  A portion is shown here 
for guidance regarding its form and content. }
\end{table}

\section{\label{sc:listdirect}List of objects found by the spectral index search  }
Description in Sect.~\ref{sc:direct}

\begin{table}[h!]
\caption{\label{tab.direct} Late-L and T dwarfs found by the  spectral index search }
\centering

\begin{tabular}{llrrrcccc}
\hline\hline
Alias (Ref.)    &    SpT       &    \Euclid  ID   &    RA(J2000)   &    Dec(J2000)   &    \IE    &  \YE  & \JE  & \HE   \\
    &           &      &       &      &     (mag)   &  (mag)    &  (mag)    &  (mag)   \\ 

\hline
  E511520 & T1: & -511520903274482292  & 03:24:36.5 & -27:26:53.6 & 22.9 & 19.6 & 19.1 & 18.7\\
  E517518$^4$ & L9 & -517518361295768184  & 03:27:00.4 & -29:34:36.5 & 23.3 & 19.9 & 19.3 & 18.8\\
  E523574 & T4 & -523574860290315045  & 03:29:25.8 & -29:01:53.4 & 26.7 & 21.4 & 20.9 & 21.1\\
  E528241 & T5 & -528241075263163744  & 03:31:17.8 & -26:18:58.9 & 27.6 & 23.6 & 23.2 & 23.6\\
  E536416 & T2: & -536416224285940430  & 03:34:34.0 & -28:35:38.6 & 25.4 & 21.5 & 21.2 & 20.9\\
  E581332$^1$ & T7 & -581332495491830038  & 03:52:32.0 & -49:10:58.8 & 24.3 & 20.0 & 19.5 & 20.1\\
  E597913$^1$ & T7 & -597913643476826162  & 03:59:09.9 & -47:40:57.4 & 24.8 & 20.2 & 19.8 & 20.2\\
  E644720$^4$ & T1: & -644720877461587627  & 04:17:53.3 & -46:09:31.5 & 22.9 & 19.5 & 19.2 & 18.9\\
  E265716$^4$ & T4p & 2657163304658383990  & 17:42:51.9 & +65:50:18.2 & 24.4 & 20.5 & 20.1 & 19.8\\
  E266485$^2$ & T6: & 2664850113649936423  & 17:45:56.4 & +64:59:37.1 & 24.7 & 20.4 & 20.0 & 20.5\\
  E267056$^4$ & L9: & 2670569747654000953  & 17:48:13.7 & +65:24:00.3 & 24.0 & 20.7 & 20.0 & 19.4\\
  E271006$^3$ & T2 & 2710066793674540980  & 18:04:01.6 & +67:27:14.8 & 24.6 & 20.3 & 19.9 & 19.5\\
  E271934$^{4,\,  5}$ & L9p & 2719340730667146696  & 18:07:44.2 & +66:42:52.8 & 23.8 & 20.4 & 19.9 & 19.5\\
  E273015$^4$ & T1: & 2730150213677979458  & 18:12:03.6 & +67:47:52.6 & 23.2 & 20.2 & 19.8 & 19.7\\
  E273062 & T3p & 2730620775659672177  & 18:12:14.9 & +65:58:02.0 & 24.0 & 20.9 & 20.6 & 20.5\\
  E274809$^3$ & T0 & 2748094058670347269  & 18:19:14.3 & +67:02:05.0 & 23.3 & 19.4 & 19.0 & 18.7\\
\hline
\end{tabular}
\tablecomments{ The object alias with no notes are discoveries by the spectral index search,  the rest are: 
(1) two already cited in Zhang's compilation (and references therein); 
(2) one cited in  \cite{Mace2013}; 
(3) two already found  by {\v Z}erjal et al. (in prep.); 
(4) six discovered by the spectral index search  that were later found in the process of creating {\v Z}erjal's catalog;
(5) one also included in Mohandasan et al. (in prep.). 
Uncertainties in the spectral classification $> 1$ subtype are noted by a colon; peculiar objects are indicated by ``p". 
The magnitudes are from  {\v Z}erjal et al. (in prep.). }

\end{table}

\newpage

\section{\label{sc.marusafull} Spectroscopically confirmed UCDs from {\v Z}erjal's catalog}
Description in Sect.~\ref{sc:Marusaspec}

\begin{table}[h!]
\digitalasset
\caption{\label{tab.marusa} Descriptive version of the ``Spectroscopically confirmed  ultracool   dwarfs from {\v Z}erjal's catalog" table.  }
\centering

\begin{tabular}{llrrr}
\hline\hline
Alias    &    SpT       &    \Euclid  ID   &   RA(J2000)   &    Dec(J2000)        \\
\hline
  E507757 & L1 & -507757799278695877 & 03:23:06.2 & -27:52:10.5\\
  E508475 & M8 & -508475889270441632 & 03:23:23.4 & -27:02:39.0\\
  E513018 & L1 & -513018452289856025 & 03:25:12.4 & -28:59:08.2\\
  E513604 & L0 & -513604620270371688 & 03:25:26.5 & -27:02:13.8\\
  E514230 & L0 & -514230855279574983 & 03:25:41.5 & -27:57:27.0\\
  E516579 & L1: & -516579302266810335 & 03:26:37.9 & -26:40:51.7\\
  E516938 & L0p & -516938378281012036 & 03:26:46.5 & -28:06:04.3\\
  E518567 & M9 & -518567715273284821 & 03:27:25.6 & -27:19:42.5\\
  E519971 & T6 & -519971822278279190 & 03:27:59.3 & -27:49:40.5\\
  E520726 & L0 & -520726476283921213 & 03:28:17.4 & -28:23:31.6\\
  
\hline
\end{tabular}
\tablecomments{This list exclude the  objects found in Zhang's compilation and in the  spectral index search.  
Uncertainties in the spectral classification $> 1$ subtype are noted by a colon; peculiar objects are indicated by ``p". 
The table is published in its entirety in the electronic 
edition of the {\it Astrophysical Journal}.  A portion is shown here 
for guidance regarding its form and content. }
\end{table}

\section{\label{sc.appTeff} Effective temperature estimation }
Description in Sect.~\ref{sc:teff}.

\begin{table}[h!]
\digitalasset
\caption{ Descriptive version of the ``Effective temperatures ($T_{\mathrm{eff}}$)" table.}
 \label{tab.teffalll}
 \centering          
 \begin{tabular}{llr}
  \hline\hline
  Alias & SpT & $T_{\rm eff}$  \\
    & & (K)  \\
  \hline
  \noalign{\smallskip}
E581332  &   T7  &  $ 783_{-37}^{+75\phantom{0}}$ \\
E523574  &   T4  &  $ 1161_{-142}^{+38}$  \\
E271006  &   T2  &  $ 1376_{-53}^{+92\phantom{0}}$ \\
E607287  &   T0  &  $ 1440_{-146}^{+79}$  \\
E517518  &   L9  &  $ 1455_{-76}^{+39\phantom{0}}$ \\
E638441  &   L5: &  $ 1462_{-122}^{+50}$ \\
E576081  &   L5: &  $ 1480_{-129}^{+121}$ \\
E592333  &  L3: &  $ 1495_{-95}^{+85\phantom{0}}$ \\

\noalign{\smallskip}                              
  \hline
 \end{tabular}
 \tablecomments{
 Uncertainties in the spectral classification $> 1$ subtype are noted by a colon; peculiar objects are indicated by ``p". 
 This table is published in its entirety in the electronic 
edition of the {\it Astrophysical Journal}.  A portion is shown here 
for guidance regarding its form and content. }
\end{table}

\bibliography{all}{}
\bibliographystyle{aasjournal}



\end{document}